%% file: main.tex
\documentclass[12pt]{article}

\usepackage{PRIMEarxiv}

\usepackage[utf8]{inputenc} % allow utf-8 input
\usepackage[T1]{fontenc}    % use 8-bit T1 fonts
\usepackage[hidelinks]{hyperref}       % hyperlinks
\usepackage{url}            % simple URL typesetting
\usepackage{booktabs}       % professional-quality tables
\usepackage{amsfonts}       % blackboard math symbols
\usepackage{nicefrac}       % compact symbols for 1/2, etc.
\usepackage{microtype}      % microtypography
\usepackage{lipsum}
\usepackage{fancyhdr}       % header
\usepackage{graphicx}       % graphics
\graphicspath{{media/}}     % organize your images and other figures under media/ folder

\usepackage{amsthm} % for proof and theorem environments
\usepackage{amsmath} % for math text
\usepackage{xcolor}
\newcommand{\nicole}[1]{{\color{black}  #1}} %blue
\newcommand{\nabin}[1]{{\color{black}  #1}} %red
\usepackage{setspace}
\newtheorem{theorem}{Theorem}[section]

\usepackage{titlesec}

\titlespacing*{\section}{0pt}{0pt plus 0pt minus 0pt}{0pt plus 0pt minus 2pt}
\titlespacing*{\subsection}{0pt}{0pt plus 0pt minus 0pt}{0pt plus 0pt minus 2pt}
\titlespacing*{\subsubsection}{0pt}{0pt plus 0pt minus 0pt}{0pt plus 0pt minus 1pt}

\usepackage{graphicx}
\usepackage{subcaption}

\title{Bayesian Sparsity Modeling of Shared Neural
Response in Functional Magnetic Resonance Imaging
Data
}

{\small
\author{
  Spencer Wadsworth \\
  Department of Statistics \\
  University of Connecticut \\
  Storrs, CT, USA\\
  \texttt{wadsworth.spencer.g@gmail.com} \\
   \And
  Nabin Koirala \\
  Brain Imaging Research Core \\
  University of Connecticut \\
  Storrs, CT, USA\\
  \texttt{nabin.koirala@uconn.edu} \\
  \\
  \And 
  Nicole Landi \\
  Department of Psychological Sciences \\
  University of Connecticut \\
  Storrs, CT, USA\\
  \texttt{nicole.landi@uconn.edu} \\
  \And
  Ofer Harel \\
  Department of Statistics \\
  University of Connecticut \\
  Storrs, CT, USA\\
  \texttt{ofer.harel@uconn.edu} \\
}

}

\usepackage[pagewise]{lineno}
\usepackage{natbib}
\begin{document}
\maketitle
\thispagestyle{empty}
% \doublespacing
% \setstretch{2.3}
\begin{abstract}
Detecting shared neural activity from functional magnetic resonance imaging (fMRI) across individuals exposed to the same stimulus can reveal synchronous brain responses, functional roles of regions, and potential clinical biomarkers. Intersubject correlation (ISC) is the main method for identifying voxelwise shared responses and per-subject variability, but it relies on heavy data summarization and thousands of regional tests, leading to poor uncertainty quantification and multiple testing issues. ISC also does not directly estimate a shared neural response (SNR) function.
We propose a model-based alternative applicable to both task-based and naturalistic fMRI that simultaneously identifies spatial regions of shared activity and estimates the SNR function. The model combines sparse Gaussian process estimation of the response function with a Bayesian sparsity prior inspired by the horseshoe prior to detect voxel activation. A spatially structured extension encourages neighboring voxels to exhibit similar activation patterns. We examine the model’s properties, evaluate performance via simulations, and analyze two real-world fMRI datasets, including one task-based and one naturalistic dataset.
The Bayesian framework provides principled uncertainty quantification for the shared response function and shows improved activation detection and response estimation compared to standard approaches. Model fits demonstrate comparable or superior performance relative to ISC, while the framework opens avenues for clinical applications.
\end{abstract}

\keywords{Sparse Gaussian processes \and
          Shrinkage priors \and
          Spatial statistics \and
          Intersubject correlation \and
          Neuroimaging \and 
          Naturalistic fMRI 
          }

\clearpage
\pagenumbering{arabic}
\setcounter{page}{1}
\setstretch{1.4}
% \doublespacing
\normalsize
% \linenumbers
\input{manuscript_rev}

%Bibliography
% \bibliographystyle{unsrt}  
\bibliographystyle{plainnat}
\bibliography{references} 

\end{document}

\maketitle
\section*{Web Appendix A. Proof of Theorem 1 and Figure of Marginal HTHS Prior}
\begin{proof}
Without loss of generality, assume $\phi = 1$ and drop it to simplify 
notation. Then,

\[
        p(\beta) = \int_0^{\infty}\frac{1}{\sqrt{2 \pi \lambda^2}} \text{exp} \left( -\frac{\beta^2}{2\lambda^2}\right) \frac{\Gamma(\frac{\nu + 1}{2})}{\sqrt{\nu \pi}\Gamma(\frac{\nu}{2})} 
       \left(\frac{\nu + \lambda^2}{\nu} \right)^{-(\frac{\nu + 1}{2})} d\lambda 
\]        
\\~\\
Let $\zeta = 1/ \lambda^2$, then
\[       
    \begin{aligned}
       p(\beta) &= R\int_0^{\infty} \frac{1}{\zeta} \text{exp} \left( -\frac{\beta^2}{2}\zeta\right) \left(\frac{\nu + (1/\zeta)}{\nu} \right)^{-(\frac{\nu + 1}{2})} d\zeta \\\\
        &\geq R\int_0^{\infty} \frac{1}{\zeta} \text{exp} \left( -\frac{\beta^2}{2} \zeta\right) \text{exp} \left( -\frac{(\nu + 1)}{2 \nu} 
        \frac{1}{\zeta} \right) d\zeta\\\\
        &\geq R\int_0^{\infty} \frac{1}{\zeta} \text{exp} \left( -\frac{\beta^2}{2} \zeta \right) \text{exp} \left( - 
        \frac{1}{\zeta} \right) d\zeta\\\\
        &\geq R\int_1^{\infty} \frac{1}{\zeta} \text{exp} \left( -\frac{\beta^2}{2} \zeta \right) d\zeta \\\\
        &= R E_1\left(\frac{\beta^2}{2}\right)
    \end{aligned}
\]
\\~\\
where $E_1(\cdot)$ is the exponential integral function. 
The first inequality follows from the fact that $(1 + x)^{-a} \geq e^{-ax}$ and the second from the fact that $(\nu + 1)/2\nu \geq 1$ for 
$\nu \geq 1$. 
The function $E_1(\cdot)$ has the lower bound\\~\\
\[
  \frac{\text{exp}(-t)}{2} \text{log} \left(1 + \frac{2}{t}\right) <
  E_1(t) 
\] 
\\~\\
for all $t > 0$. From this bound $p(\beta) \rightarrow \infty$
as $\beta \rightarrow 0$.

\end{proof}

\begin{figure*}[!hbt]
    \centering
    \includegraphics[scale=.55]{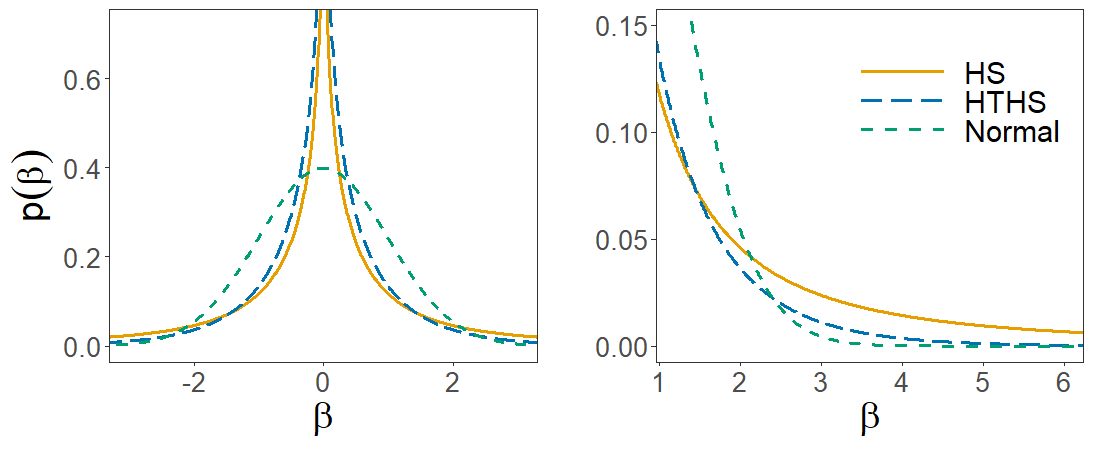}
    \caption{Marginal horseshoe, half-t horseshoe, and 
    standard normal prior distributions. Here $\tau = 1$ and 
    $\phi = 1$.}
    \label{fig:marg_priors}
\end{figure*}

Figure \ref{fig:marg_priors} compares the marginal 
HTHS with $\nu = 1000$ with
the marginal HS prior where $\tau = \phi = 1$ and 
with the standard normal distribution. The figure shows 
plots for bulk of the prior distributions near zero as well 
as the distribution tails.

\section*{Web Appendix B. Shrinkage Factor Estimates Under Differing Prior Distributions}

\begin{figure}[!hbt]
    \centering
    \includegraphics[width=1\linewidth]{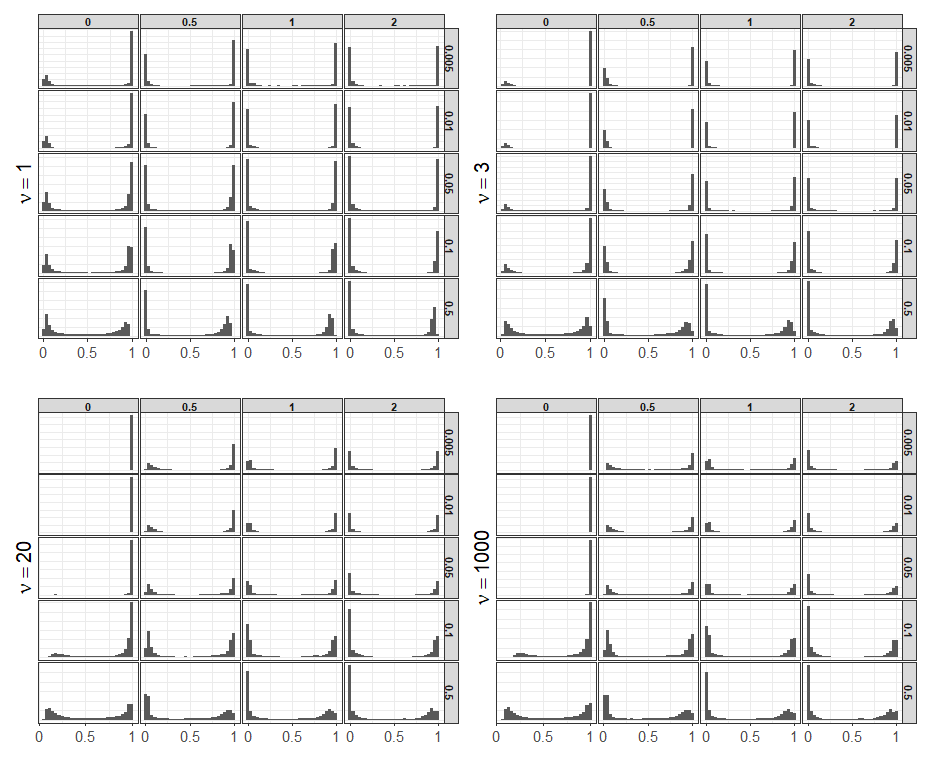}
    \caption{Histograms of posterior mean of shrinkage factor $\kappa_v$ 
    from simulation study in Section 4 of the main text. Each of the four
    plots represents results where for different degrees of freedom
    $\nu$ of the HTHS prior local shrinkage parameter. Each plot is
    further faceted by the shared neural response function 
    signal strength on the x-axis and the prior parameter $\tau^*$
    of the prior assigned to the HTHS 
    global shrinkage parameter.}
    \label{fig:shrink_hist}
\end{figure}

Figure \ref{fig:shrink_hist} shows histograms of the estimated posterior
mean of the shrinkage factor $\kappa_v$ from equation (8) in the main
text. These pertain to the simulation study in Section 3. There are four 
plots, one for each $\nu \in \{1, 3, 20, 1000\}$ where $\nu$ is the 
degrees of freedom of the half-t distribution in the HTHS prior. 
Each plot contains subplots faceted by activated signal strength in the 
x-axis and the hyperprior parameter value $\tau^*$ in the y-axis.
Recall from Section 2 that the prior distribution assigned to
the global shrinkage parameter of the HTHS prior is 
$\tau \sim t^+_{1000}(0, \tau_0)$ where 
$\tau_0 = \tau^*/\sqrt{S\times V \times T}$. Here $\tau^*$ is considered
the rough prior estimate of the
proportion of active voxels in a naturalistic SNR
experiment.

The plots
show the characteristic U shape of the shrinkage factor, and how it varies
depending on the choice of prior. As discussed in the main text, where
the true activation signal is 0, the HTHS priors for smaller $\nu$ are 
more likely to falsely declare some voxels active. This may also occur
occasionally for larger $\nu$, but only as $\tau^*$ increases. In general,
as $\tau^*$ increases, the number of small $\kappa_v$ also increases. 
Ultimately the value $\tau^* = 0.1$ was chosen for the results presented 
in the main text because it most accurately reflects our prior beliefs 
on voxel activation, and because it maintained relatively strong 
shrinkage based on the results shown here.

\section*{Web Appendix C. Checkerboard Stimulus Activation Maps for BNR Models}

\begin{figure}[!hbt]
    \centering
    \includegraphics[width=0.55\linewidth]{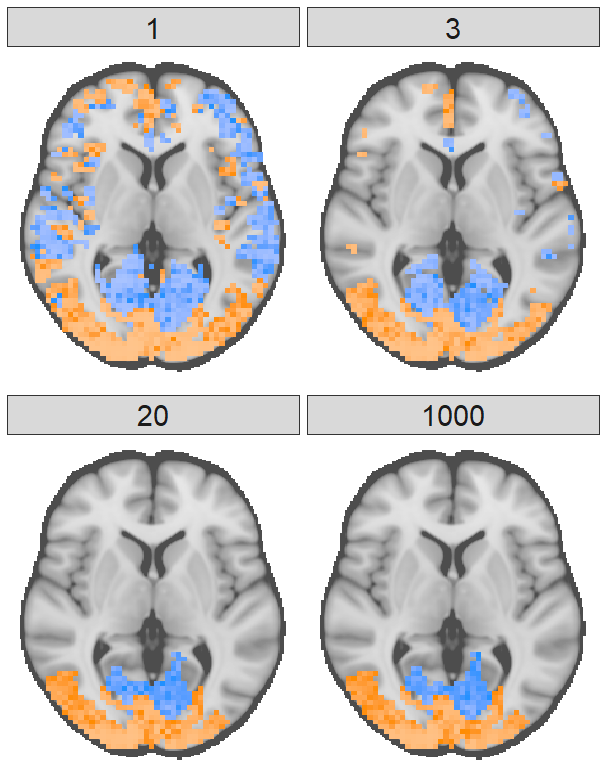}
    \caption{Shared neural activation maps of checkerboard task fMRI
    data for BNR models faceted by degrees of 
    freedom $\nu \in \{1, 3, 20, 1000\}$.}
    \label{fig:check_act_all}
\end{figure}

Figure \ref{fig:check_act_all} shows activation maps for the BNR 
SNR models for degrees of freedom $\nu \in \{1, 3, 20, 1000\}$ 
using checkerboard tfMRI data (see section 3.2 in the main text).
The purpose for including these results is to show how activation
differs by the degrees of freedom assigned to the prior of the 
local shrinkage parameter in the HTHS prior distribution. For
small
degrees of freedom, 1 and 3, the BNR model shows far more activation
than even the GLM shows.
This suggests that 
either there is far more SNR activation by the checkerboard stimulus
than would generally be detected via the GLM, or more
 likely that the BNR-1 and
BNR-3 models' activation predictions are spurious. This further 
illustrates the need to use the HTHS prior over the HS prior with
high degrees of freedom in the local shrinkage parameter prior.

% \bibliographystyle{unsrt}  
% \bibliography{references}  

%% file: manuscript_rev.tex
\section{Introduction}

% \section{Shared Neural Response and Experimental fMRI Data}

% Functional magnetic resonance imaging (fMRI)
% is a measure of the 
% blood oxygenation level dependent (BOLD) brain signals.
% fMRI is a non-invasive indirect measure of neural 
% activity,
% and corresponds with blood flow through the 
% brain as it reacts to the performance of tasks or 
% exposure to stimuli. A brain region which is activated is
% characterized by a subsequent increase of oxygen followed
% by a slight decrease. The BOLD response at activation 
% follows the hemodynamic response function (HRF).

% Written by Nicole
Intersubject neural synchronization (INS) assumes that
people who are exposed to the 
same external stimuli will exhibit similar neural activity in 
regions effected by the stimuli. A primary method
of data collection to study INS is functional 
magnetic resonance imaging (fMRI)  \citep{nastase2019measuring,jaaskelainen2008inter}.
% ,hasson2004intersubject}).
\nicole{
fMRI utilizes the blood-oxygen-level-dependent (BOLD) signal, a non-invasive and indirect measure of neural activity that reflects changes in blood flow and oxygenation in response to task performance or stimuli. 
% The BOLD signal arises from the hemodynamic response, a physiological process in which neural activity leads to increased cerebral blood flow that overcompensates for local oxygen consumption. This results in a relative increase in oxygenated hemoglobin and a decrease in deoxygenated hemoglobin. The temporal dynamics of this process, characterized by an initial rise, peak, and gradual return to baseline, often with a slight undershoot, are described by the hemodynamic response function (HRF), which is commonly used to model the BOLD signal.
} % End of Nicole
An fMRI image at a single time point consists of tens to 
hundreds of thousands of BOLD measurements in three dimensional 
cubes
known as voxels. A sequence of dozens to thousands of 
images taken of a participant's brain over the course of 
several minutes to hours of scanning produces a time series
at each voxel.
% make the data 
% functional. 
% The exact specifications of voxel volume
% or frequency of sequential imaging, known as repetition
% time (TR) depend on the capabilities of the scanning 
% machine and the aim of the research. 
% \nicole{
% Key acquisition parameters, such as voxel size (spatial resolution) and the temporal sampling rate, defined by the repetition time (TR), depend on the scanner capabilities and the goals of the study.}
See \citet{lindquist2008statistical} for a brief description of fMRI
data meant
for statisticians.

% principle scientific objective of analyzing fMRI 
% data is to study 
% \nicole{One recently popularized approach for analyzing fMRI data is
% intersubject neural synchronization (INS). 
% % or the notion that
% This method is based on the assumption that}
% regional brain activity will be similar among multiple 
% persons exposed to the same stimuli. 
% The methods used for inferential analysis of 
% neural synchronization with fMRI data 
% are dependent on the number of subjects in a study and
% the size of the data, but even more so they depend
% on the experiment imposed on a subject while undergoing
% scanning. 
The most common type of experiment in fMRI 
studies is task-based fMRI (tfMRI). In tfMRI studies, participants
are scanned while exposed to highly controlled 
simple stimuli and/or asked to perform simple
actions.
% for example viewing a blinking light or flipping a switch.  
% stimuli/actions are known as tasks and they are presented
% to participants on a prespecified timescale. 
Task-based analyses are helpful for determining neural 
reactions to basic events, and there is an abundance
of literature on analyzing tfMRI data in neuroscience
and statistical literature \citep{friston1994analysis,lindquist2008statistical,zhang2015bayesian}.
In daily life it is unlikely for people to be 
exposed to simple tfMRI tasks in isolation,
\nicole{limiting the ecological validity of such studies}
% making the such experiments of limited use 
\citep{sonkusare2019naturalistic,simony2020analysis}.
% Likely the next most common fMRI experiment is 
% resting-state fMRI (rsfMRI). 
% \nicole{Another common approach is resting-state fMRI (rfMRI),
% during which participants rest with their eyes open or closed, sometimes with a limited visual display such as a fixation cross}.
% In rsfMRI studies, participants 
% are not exposed to any stimuli during scanning but rather
% are asked to simply rest during scanning. 
% There is also
% an abundance of rsfMRI research in the statistics 
% literature.
% (see for review \cite{lindquist2025statistics}). 
% but its use in studying INS 
% is limited by lack of shared common experience,
% % which 
% % stimuli provide. 
% given no specific stimuli. 
% However, we focus this work on
% INS, so little more will be said on rsfMRI.
% A third popular
% experiment type 
% is 
\nicole{Increasingly, cognitive neuroscientists who study
INS are utilizing}
naturalistic fMRI (nfMRI), or more ecologically valid continuous 
stimuli such as listening to stories, or viewing a movie.
% In an nfMRI experiment, participants are exposed to a more 
% natural stimulus such as listening to stories, music, 
% or watching a movie. Naturalistic stimuli are 
% considered more ecologically valid 
Naturalistic stimuli can thus lead 
to insights about more complex neural processing and 
inter-person synchrony than task-based stimuli 
\citep{schmuckler2001ecological,hasson2004intersubject,nastase2020keep,simony2020analysis}, while still providing a cognitive focus which is 
absent in rsfMRI experiments \citep{saarimaki2021naturalistic,zhang2021naturalistic}.
Interestingly, while nfMRI studies 
are prevalent in the neuroscience literature and 
other fields, they have received
considerably less attention in the 
statistics literature than tfMRI despite the
growing need for advanced methodological development. 
% The study of naturalistic data promises to provide insights
% into complex physiological, behavior, and emotional 
% neural responses which are largely out of reach in the
% oversimplication of reality inherent in task-based stimuli
% and the overgeneralization of reality from 
% resting-state experiments
% \cite{saarimaki2021naturalistic,zhang2021naturalistic}.

\nicole{
% The motivation for this work is to develop robust statistical methodology for studying INS under paradigms involving shared naturalistic stimuli, while building a unified framework that can incorporate both tfMRI and nfMRI analyses. 
The tightly controlled experiments in tfMRI allow 
for explicitly constructing predictor variables of the
expected BOLD response to be used in a general linear
model (GLM), but 
% In tfMRI, experimental paradigms are tightly controlled, with stimuli or tasks presented according to a predefined design. This allows researchers to construct explicit predictor variables for the expected BOLD response, typically by convolving task timings with the hemodynamic response function (HRF), and to apply statistical modeling approaches such as the general linear model (GLM) to identify task-related brain activity. 
% However, 
this framework relies on strong assumptions about the timing and form of neural responses, likely violated in naturalistic settings. 
The processes in naturalistic stimuli which influence cognitive function cannot be easily decoded into a complete set of known predictors. Consequently, standard model-based approaches are insufficient, motivating the use of data-driven methods such as intersubject correlation (ISC), independent component analysis (ICA), and other multivariate INS techniques
\citep{nastase2019measuring, pamilo2012functional}.
}
\nabin{While current approaches such as ISC are effective for detecting time-locked similarity across participants,
they rely on heavily summarized data which ignores much
of the uncertainty in fMRI data,
they do not directly estimate the latent shared neural response, and they typically rely on large numbers of regional tests \citep{nastase2019measuring}.
Functional alignment methods, including the shared response model (SRM), move closer to latent-response estimation \citep{chen2015reduced}, 
but they rely on restrictive assumptions such as exact temporal correspondence, low-dimensional linear structure, orthogonality constraints, and simplified noise models. }

The contribution of this manuscript is to present a 
unified Bayesian 
modeling framework for analyzing INS under 
nfMRI experiments.
\nabin{The model, termed the Bayesian neural response (BNR) model, provides a joint inferential framework for estimating the latent shared response} as well as regional activation. It
improves upon the standard 
methodology currently used for analysis of nfMRI data
by
making use of all available BOLD response data, thus accounting
for much of the uncertainty ignored using ISC methods.
% The modeling framework
% also
% unifies the methodology used for analysis of tfMRI and nfMRI data. 
The BNR model is a
generalization of the classical GLM used 
in traditional 
fMRI analyses in which the expected BOLD predictor
is estimated to accommodate nfMRI experiment data.
% in (\ref{eq:bold_glm}) where the
% function representing the SNR $h_k(\cdot)$ is estimated rather than 
% specified in the design.  
Shared voxelwise estimation is done using a novel 
spatial sparsity prior.
% similar
% to the horseshoe prior introduced by Carvahlo et al. 
% \cite{carvalho2009handling, carvalho2010horseshoe} but allowed to vary in
% shape and shrinkage strength. Spatial information is also incorporated
% into the prior distribution via a sparse Gaussian Markov random field 
% prior distribution model.
% The simultaneous
% estimation of the activation coefficient along 
% with the SNR function leads to a
% lack of identifiability in model parameters.
% This necessitates the careful specification of the
% sparsity prior shape and the selection of hyperparameters. 
% A latent spatial component enforces shrinkage and thus 
% shared activation to be similar among neighboring voxels.
% The use of the sparsity prior for detecting activation provides more power in
% detecting weak and moderate SNR signals, and computationally is much easier
% to implement than spike-and-slab priors.
% The decision to use a sparse rather than a standard GP made for efficiently
% handling the large data 
% inherent in a multi-subject fMRI dataset. The prior distribution
% on the SNR allows estimating a trajectory of $h(\cdot)$ which may not be
% captured in a specified BOLD predictor, and it allows for estimating a 
% complete function which may capture the complete response to 
% naturalistic stimuli.
The motivating data come from \citet{telesford2023open} 
who provide well curated 
nfMRI as well as tfMRI response data on 22 adult 
participants. 
% The dataset contains
% multiple participants' fMRI responses to both
% naturalistic and task-based experiments which 
% is ideal for implementing
% and analyzing the BNR model. 

% The contribution of this manuscript is a Bayesian model based framework for 
% detecting
% shared neural activation and estimating the SNR function of nfMRI experiment
% data. 

Section \ref{sec:background} includes background on 
shared neural response (SNR) theory and classical GLM and ISC
methodology used for analysis of tfMRI and nfMRI data respectively.
In Section \ref{sec:bayes_mod}, the BNR model is introduced including
the nearest neighbor sparse GP used for estimation of the expected BOLD response and the 
spatial sparsity prior used for detecting shared voxel activation. 
In Section \ref{sec:analyses}, the model is fit to data in three 
examples including simulated, real tfMRI, 
and real nfMRI data.  In these analyses, the BNR model shows results
comparable to 
the classical multi-participant GLM when fit to tfMRI data and 
superior activation and function estimation compared with 
ISC methods in the nfMRI setting. The manuscript is concluded 
with a brief discussion in Section \ref{sec:conclusion}.

\section{Shared Neural Response Background} \label{sec:background}

In this section, we briefly describe SNR
theory, the largely accepted neuroscience theory for
INS. 
We also discuss the classical GLM and basic 
ISC analyses,
including some of their 
drawbacks before introducing the proposed unifying 
approach under the BNR framework. 
\subsection{Shared Neural Response Theory}
Following the notation used in
\citet{nastase2019measuring}, equation (\ref{eq:snr_eq}) 
presents a formula of the signals that make up the time series in one voxel 
for several participants exposed to the same time-locked stimulus. 
The signal $c(\cdot)$ is 
triggered by the stimulus which is shared by participant or subject
$s$ and all other participants. 
% It may reflect synchrony of 
% low-level brain processing sensory features 
% as well as high-level features such as emotional or social processing
% \cite{saarimaki2021naturalistic}.
The signal $id(\cdot)$, 
known as the idiosyncratic response, 
represents the triggered response
specific to participant $s$,
% based, for example, on a personal interpretation
of the stimulus, 
and $\epsilon(t)$ is participant specific spontaneous noise.
% Additional signals may also be added to (\ref{eq:snr_eq})
% such as those representing
% group or culture specific responses.
% Figure \ref{fig:snr_nastase}, taken from Nastase et al.,
% shows how the three signals compare between two participants $A$ and $B$.
\begin{equation}
    x_s(t) = \beta_s c(t) + \alpha_s id_s(t) + \epsilon_s(t)
    \label{eq:snr_eq}
\end{equation}

% The underlying assumption about $c(\cdot)$ is that the 
% pairwise correlation of this signal between participants is equal to or
% approaching 1. 
% Another assumption of the shared signal is that as the total number 
% of participants $S$ in a
% study increases, the mean of the overall signals at time $t$, 
% $x_s(t)$, approaches $c(t)$, that is 
% $\lim_{S \to \infty} \bar{x}(t) = c(t)$ where 
% $ \bar{x}(t) = \sum_{s = 1}^S x_s(t)$.
Inference focuses on estimating  
$c(\cdot)$ and $\beta_s$. 
The form of $c(\cdot)$ should correspond with the stimuli, and it is estimated from the data or given in a model.
The parameter $\beta_s$ reflects how strong $c(\cdot)$
is, and a significant value gives evidence of the presence of INS.
% A common presentation of analysis results is to report a statistical
% parametric/activation map, or an image of a brain with voxels highlighted where
% shared voxel activation occurs in response to a stimulus
% (see Figures \ref{fig:check_act} and \ref{fig:dme_act} for examples).

% \begin{figure*}[!hbt]
%     \centering
%     \includegraphics[width=0.7\linewidth]{images/nastase_snr_brains.jpeg}
%     \caption{Figure showing logic of shared neural response.
%     For two brains signals recorded from persons participants
%     with the same 
%     time locked stimuli, the BOLD signal at corresponding voxels
%     should be made up of three components: the 
%     shared neural response driven by the stimulus, 
%     participant specific 
%     idiosyncratic signal, and participant specific spontaneous
%     noise. \\
%     Figure reproduced from Nastase et al.\ \cite{nastase2019measuring}, 
% licensed under the Creative Commons Attribution 4.0 International 
% License (CC BY 4.0, http://creativecommons.org/licenses/by/4.0/).}
%     \label{fig:snr_nastase}
% \end{figure*}

\subsection{Classical GLM}
% One of the most common neural experiments conducted and 
% analyzed with fMRI data is a task-related study. In task
% fMRI experiments, during an fMRI scanning session, a participant
% is exposed to one or more simple tasks. 
The most common method for analyzing stimulus driven fMRI data is
the classical GLM, first 
proposed by \citet{friston1994analysis},
see also  
\citet{lindquist2008statistical}.
The tfMRI data is assumed to follow model
(\ref{eq:bold_glm}) where $B_{s,v}(t)$ is the BOLD response
at time $t$ for participant $s$ and voxel $v$. 
Here, there are $K$ separate tasks, and $h_k(\cdot)$ is the task BOLD predictor. 
The term $\epsilon_{s,v}(t)$ captures noise and possible nuisance 
variables.

\begin{equation}
    B_{s,v}(t) = \sum_{k = 1}^K\beta_{k,s,v}h_{k}(t) + \epsilon_{s,v}(t)
    \label{eq:bold_glm}
\end{equation}

\begin{figure*}[!hbt]
    \centering
    \includegraphics[width=0.75\linewidth]{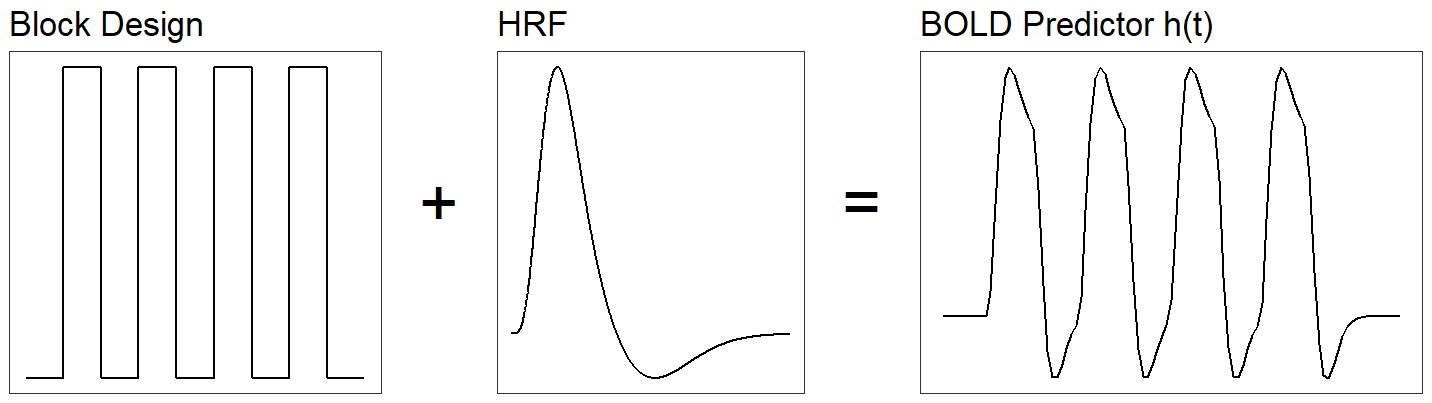}
    \caption{Process of defining the BOLD predictor in 
    the classical task-based GLM}
    \label{fig:bold_pred}
\end{figure*}

The process of defining the task BOLD predictor is shown in
Figure \ref{fig:bold_pred}. 
Here, the block design --or timing of whether a task is on or off--
is convolved with the hemodynamic response function (HRF) to produce
the BOLD predictor, see \citet{zhang2015bayesian} for additional discussion.
% The HRF
% is the assumed BOLD behavior of an activated brain 
% region immediately following a stimulus. The 
% double-gamma HRF is considered canonical, though 
% others forms have been studied (see 
% \cite{zhang2015bayesian} for examples
% ). 

The main objective of a task-based GLM is to detect voxel activation
by fitting model (\ref{eq:bold_glm}) to
the data and testing whether or not $\beta_{k,s,v}$ is zero, where 
a significant result leads to the conclusion that voxel $v$ is 
activated by task $k$.
The activation patterns are reported in an activation map, or a map of the brain where active voxels are highlighted (see Figures \ref{fig:check_act} and \ref{fig:dme_est} as examples). 
% This is typically done 
% by fitting model (\ref{eq:bold_glm}) to
% the data and testing whether or not $\beta_{k,s,v}$ is zero, where 
% a significant result leads to the conclusion that voxel $v$ is 
% activated by task $k$. 
In multi-participant studies, model (\ref{eq:bold_glm}) is fit to each
participant and voxel independently, and the
estimated coefficients are treated as data on which tests are 
conducted.
Extensions of the classical GLM deal with 
multiple testing issues, Bayesian modeling for including voxel spatial correlation, prior distributions for sparse activation,
and computational improvements
\citep{zhang2015bayesian}.

In tfMRI experiments, it is possible to formulate a reasonable
set of predictors $\{h_k(\cdot)\}_{k = 1}^K$, although the 
design may not capture completely
the BOLD trajectory driven by a stimulus.
\citet{wilzen2020physiological}
proposed modeling the BOLD response 
via a Gaussian process (GP) prior,
but their analysis is limited to a 
single participant experiment,
and they assume apriori that the mean function of the 
GP prior is an already predicted task BOLD response.
% , restricting
% somewhat the trajectory of the estimated function.
In nfMRI studies using the classical GLM, 
stimulus features are extracted either manually or using
computational methods such as neural networks
\citep{gucclu2015deep,jain2018incorporating}. However,
the complexity of naturalistic stimuli generally makes it
impossible to extract every relevant feature, and those 
obtained via computational methods can be highly correlated
with eachother \citep{eickenberg2017seeing,la2022feature}.

% must be decoded or extracted directly from the 
% continuous stimulus, but the stimulus complexity 
% generally makes it impossible to capture every relevant feature
% % one must
% % extract features from the stimuli to convolve with the HRF
% % and use as a predictor.
% % Features in naturalistic stimuli are
% % much more complicated than the simple on/off features of 
% % task stimuli, and they may change continuously throughout
% % the stimuli such as volume or contrast. 
% Features can be coded manually or extracted using 
% computational methods such as neural 
% networks \cite{gucclu2015deep,jain2018incorporating}. However, these features will 
% not contain all relevant information driving
% the BOLD response, and extracted features can be 
% highly correlated with one another 
% \cite{eickenberg2017seeing,la2022feature}.

% \[
%     y(t) = \int_0^t x_k(m)h(t - m)dm
% \]

% \begin{figure}[!hbt]
%     \centering
%     \includegraphics[scale=.55]{images/bold_convolve.png}
%     \caption{}
%     \label{fig:bold_conv}
% \end{figure}

\subsection{Intersubject Correlation}

Intersubject correlation (ISC) was introduced by
\citet{hasson2004intersubject} as a data driven method for detecting
shared activity in situations 
where model specification is difficult or incomplete,
% , such as where one is
and it has since become the standard method for analyzing the SNR 
in nfMRI data.
To perform ISC is at a given voxel, one calculates the 
pairwise correlation of the 
nfMRI time series between all participants. The result is a set of 
$S(S-1) / 2$ correlations --where $S$ is the total number of 
participants-- to be tested for significance and plotted in an 
activation map. 
Nonparametric tests performed via various methods
of permuting either raw fMRI or ISC data 
\citet{theiler1992testing,chang2020naturalistic} is the standard in ISC
analyses, but most studies ignore various levels of correlation 
including spatial correlation and/or shared participant correlation values.
\citet{chen2019integrative, chen2020untangling} 
attempted to address the lack of accounting for correlations, but this remains an open problem.
% To account for multiple testing issues, 
% nonparametric permutation tests are made with common resampling methods
% being circular shifting, phase scrambling, or bootstrap resampling 
% \cite{theiler1992testing,chang2020naturalistic}. 
% Chen et al. introduced a Bayesian model of the correlations
% which attempts to alleviate the problem of multiple testing, account
% for correlations between brain regions and participants, and allow for including
% group information in the model
% \cite{chen2019integrative, chen2020untangling}.
% In truth, the multiple testing issues are not alleviated as much as they
% are ignored, and for anymore than a handful of
% participants this model is 
% computationally prohibitive to fit using standard posterior
% sampling methods.
By construction, the ISC hypothesis tests and models 
rely on highly summarized
data and thus ignore much of the uncertainty present in the 
raw fMRI data.
In ISC literature, the SNR function is often 
estimated as the mean
of time series among participants
\citep{nastase2019measuring}, but this
lacks regularization and uncertainty quantification
and is susceptible to irregular or extreme behavior. 
% from one or a few participants. 
% Modifications of ISC exist which attempt to incorporate spatial or temporal
% relationships, such as intersubject function correlation (ISFC) or spatial 
% ISC, but these 
\section{Methodology}
\label{sec:bayes_mod}

\subsection{Shared BOLD Bayesian Response Model}

The BNR model for the shared BOLD response is given in equation
(\ref{eq:general_model}). 
% Here $B_{s,v}(t)$ is the BOLD response for participant 
% $s$ at voxel $v$ and time $t$. 
The function $f(\cdot)$ is analogous to $c(\cdot)$ in equation (\ref{eq:snr_eq}), and it captures the 
BOLD signal shared by some or all participants. 
$\beta_v$ is the 
strength of the shared signal at voxel $v$, and
$\epsilon_{s,v}(\cdot)$ captures random noise. It is possible to generalize $\epsilon_{s,v}(\cdot)$ to 
capture a participant's idiosyncratic response or
nuisance variables.
\begin{equation}
  % \begin{aligned}
    B_{s,v} (t) = \beta_{v} f(t) + \epsilon_{s,v}(t); \quad \epsilon_{s,v}(t) \sim N(0, \sigma^2_{\epsilon})
  % \end{aligned}
  \label{eq:general_model}
\end{equation}
The form of $f(\cdot)$ depends on the 
prior beliefs in a particular experiment. 
For example, in a tfMRI experiment with a single stimulus,
% if during the fMRI 
% scanning session all participants are exposed to the same single 
% task-based stimulus,
$f(\cdot) = h(\cdot)$ where $h(\cdot)$ is the 
BOLD predictor from (\ref{eq:bold_glm}). In this case 
(\ref{eq:general_model}) reduces to the classical GLM.

In the case where participants are exposed to naturalistic stimuli, 
$f(\cdot)$ can no 
longer be specified as a known function 
which captures all
aspects of the stimuli. 
% It is often desirable to estimate $f(\cdot)$,
% and
% a common strategy is to simply take the mean of the BOLD response at a specified
% voxel or region of interest over all participants \cite{nastase2019measuring}. 
% Estimating the function in this way may be reasonable, but it lacks 
% valuable regularization and can be susceptible to irregular behavior
% from just one participant such as a spike in the BOLD response.
Our desire is to use the data to simultaneously estimate the shared BOLD 
response function and the shared activation. We 
propose to embed model
(\ref{eq:general_model}) into a Bayesian framework with thoughtfully
chosen prior distributions for $f(\cdot)$ and $\beta_v$. 
The function 
$f(\cdot)$ is modeled as a latent nearest-neighbor Gaussian process 
(NNGP), reviewed in the following section. 
Importantly, estimating $f(\cdot)$ accurately depends
on having more than one participant in a study as with
only one participant, the estimated response is confounded
with the idiosyncratic signal and additional noise
in equation (\ref{eq:snr_eq}).
\subsection{Nearest-Neighbor Gaussian Process}

The NNGP was introduced in 2016 by
\citet{datta2016hierarchical,datta2016nearest}
as a sparse version of the standard Gaussian process (GP)
\citep{rasmussen2006gaussian}. 
To NNGP is defined by first constructing a standard GP. 
Let $w(t)$ be the realization of a process at time (and/or location)
$t \in \mathcal{T} \subset \mathbb{R}$. For any finite set of 
realizations $w(\boldsymbol{t}) = (w(t_1), ..., w(t_n))$ where
$\boldsymbol{t} = \{t_1, ..., t_n\}$, $w(\boldsymbol{t})$ 
follows a GP if it has
a joint Gaussian distribution, that is 
\[
  w(\boldsymbol{t}) \sim N\left(\boldsymbol{\mu}, \boldsymbol{K}(\theta
  )\right)  
\]
for some mean vector $\boldsymbol{\mu}$ and covariance matrix 
$\boldsymbol{K}(\theta)$. 
A GP is specified by its mean function 
$m_{\theta}(t) = E[w(t) | \theta]$ and its covariance function
$k_{\theta}(t, t') = \text{cov}(w(t), w(t'))$, and 
the process is denoted as
$w(t) \sim \mathcal{GP}(m_{\theta}(\cdot), k_{\theta}(t, t'))$. It is 
common when assigning a function a GP prior to set 
$m_{\theta}(\cdot) = 0$, and we follow this convention herein.

The joint density
of a GP may be written as 
$p(\boldsymbol{w}_{\mathcal{S}}) = \prod_{i = 1}^n p(w(t_i)|w(t_{1:(i-1)})$
% in equation (\ref{eq:gp_joint})
where $t_{1:i} = (t_1, ..., t_i)$.
% \begin{equation}
%   p(\boldsymbol{w}_{\mathcal{S}}) = \prod_{i = 1}^n p(w(t_i)|w(t_{1:(i-1)})
%   \label{eq:gp_joint}
% \end{equation}

% GP regression has been used extensively in modeling time series 
% \cite{brahim2001bayesian, brahim2004gaussian},
% spatial fields \cite{cressie2011statistics,cressie2015statistics}, 
% computer experiments 
% \cite{kennedy2001bayesian,gramacy2008bayesian}, and in many 
% fields of application. 

The GP has been used extensively in many fields
\citep{brahim2004gaussian,cressie2015statistics,gramacy2008bayesian} and maintains popularity largely due to its accuracy 
in making out-of-sample predictions and for the simplicity in 
interpolating unobserved points in stochastic fields with automatic
uncertainty quantification \citep{datta2016nearest}. 
% due to
% well known properties of the multivariate Gaussian distribution. 
However, evaluating the 
multivariate Gaussian likelihood becomes the GPs primary bottleneck
as one must calculate the 
determinant or inverse of the $n \times n$ covariance matrix 
$\boldsymbol{K}(\theta)$ which costs roughly $\mathcal{O}(n^3)$ floating point
operations (flops). 
% Much work has been done to ease the computational burden
% of fitting a GP by formulating approximate GPs, see Heaton et al.
% for a study comparing methods for 
% approximating GPs \cite{heaton2019case}. 

The NNGP is a sparse precision GP
approximation, and it is an example of a Vecchia approximation 
\citep{vecchia1988estimation, katzfuss2021general}. It is 
constructed by using the joint distribution
defined in (\ref{eq:gp_joint_vec}) where 
$N(t_i)$ is the set of nearest neighbors that $w(t_i)$ is conditioned on, and its cardinality is the selected 
number of neighbors $m$.
Here $\mathcal{S}$ is a set of nodes and 
$N_{\mathcal{S}} = \{N(t_i ); i = 1, 2, .., n \}$ is the collection of 
condition sets. The pair 
$\{\mathcal{S}, N_{\mathcal{S}}\}$ is a directed graph where 
$N(t_i)$ is 
the set of directed neighbors of $t_i$, that is every node in $N(t_i)$ has 
a directed edge to $t_i$.
\begin{equation}
  \tilde{p}(\boldsymbol{w}_{\mathcal{S}}) = \prod_{i = 1}^n p(w(t_i)|w(t_{N(t_i)})
  \label{eq:gp_joint_vec}
\end{equation}
\citet{datta2016hierarchical} 
show that this construction formulates a Gaussian process with 
a sparse covariance matrix $\tilde{\boldsymbol{C}}_{\theta}$ which has at most
$km(m + 1)q^2/2$ nonzero entries, where $k$ is the 
number of unobserved points in the process which one
wishes to estimate and $m$ is the selected number of neighbors. 
The required computation is then
reduced to $\mathcal{O}((n + k)m^2)$ flops.
% flops as compared to $\mathcal{O}(n^3)$
% flops for the full GP
% (see \cite{datta2016hierarchical} and 
% appendices therein for further details).

% Viewing $\mathcal{S}$ as nodes and 
% $N_{\mathcal{S}} = \{N(t_i ); i = 1, 2, .., n \}$ as the collection of 
% condition sets, the pair 
% $\{\mathcal{S}, N_{\mathcal{S}}\}$ can be viewed as a directed graph where 
% $N(t_i)$ is 
% the set of directed neighbors of $t_i$. That is, every node in $N(t_i)$ has 
% a directed edge to $t_i$. A chain of nodes 
% $t_{i_1}, t_{i_2}, ..., t_{i_z}$ where $t_{i_1} = t_{i_z}$ in a directed graph is 
% said to have a directed cycle if there is a direct edge between every
% $t_{i_j}$ and $t_{i_{j + 1}}$. If there is no directed cycle, the directed graph
% is a direct acyclic graph. If $\{\mathcal{S}, N_{\mathcal{S}}\}$ is a 
% direct acyclic graph, then the product in (\ref{eq:gp_joint_vec}) is a proper
% multivariate joint density. 

% Properties of the multivariate Gaussian distribution show that each 
% conditional distribution in (\ref{eq:gp_joint_vec}) takes the Gaussian form
% $p(w(t_i)|w(t_{N(t_i)}) = N(w(t_i)| \boldsymbol{B}_{t_i} w(t_{N(t_i)}, \boldsymbol{F}_{t_i})$
% where 
% $\boldsymbol{B}_{t_i} = \boldsymbol{C}_{t_i, N(t_i)} \boldsymbol{C}^{-1}_{N(t_i)}$
% and 
% $\boldsymbol{F}_{t_i} = \boldsymbol{C}(t_i, t_i) - \boldsymbol{C}_{t_i, N(t_i)} \boldsymbol{C}^{-1}_{N(t_i)}\boldsymbol{C}_{N(t_i), t_i}$.

% \begin{equation}
%   \tilde{p}(\boldsymbol{w}_{\mathcal{S}}) = 
%   \prod_{i = 1}^n N(w(t_i)| \boldsymbol{B_{t_i}} w(t_{N(t_i)})
%   \label{eq:gp_vec_norm}
% \end{equation}

We assign an NNGP prior distribution
to the latent SNR function 
$f(\cdot)$ in model (\ref{eq:general_model}),
that is $f(t) \sim NNGP(0, k_{\theta}(\cdot, \cdot))$ where the covariance
function is set to be the squared exponential function in 
(\ref{eq:sq_exp}). Here, $\sigma^2$ determines
the variance and $\rho$ is known as the lengthscale parameter which controls
the distance at which the correlation between points $x$ and $x'$ decays. 
\begin{equation}
  k(x, x') = \sigma^2 \text{exp} \left( -\frac{(x - x')^2}{2 \rho^2} \right)
  \label{eq:sq_exp}
\end{equation}
It should be noted that $\beta_v$ and $f(\cdot)$ in 
(\ref{eq:general_model}) are not identifiable 
parameters, so careful
prior selection and certain constraints are important
for making estimation possible.
To assist with identifiability, the variance parameter in (\ref{eq:sq_exp}) is fixed to be $\sigma^2 = 1$ to allow $\beta_v$ from model
(\ref{eq:general_model}) to control the process variance. 
The prior assigned to $\beta_v$ is 
restricted to have nonnegative support, and additional features
of its design further encourage identifiability.
The 
lengthscale parameter $\rho$ is estimated via posterior updating
after assigning 
a prior distribution.

\subsection{Spatial Sparsity Activation Prior}

The parameter $\beta_v$ in (\ref{eq:general_model}) is typically
modeled by voxel and subject independently, or a spatial field 
is applied so that it closely resembles its neighbors 
\citep{zhang2015bayesian}.
% determines the strength
% of the BOLD SNR signal at voxel $v$ shared by the participants, and 
% In a 
% single participant tfMRI analysis,
% a nonzero $\beta_v$ indicates that a particular task leads to activity in 
% voxel $v$.
% Thus in a task-based GLM analysis, one is interested not only in the
% magnitude of $\beta_v$ but also whether or not it is equal to 0. The latter is 
% usually determined via a hypothesis test 
% \cite{friston1994statistical,friston1995analysis}. 
% in much of the task-based
% literature is modeled by participants and voxels independently.
% This naturally leads to multiple testing issues which are often resolved
% using corrections, nonparametric methods, or random field 
% theory
% \cite{nichols2002nonparametric, nichols2003controlling}. 
% Applying a spatial field to $\beta_v$ so that it closely resembles its 
% neighbors has gained some popularity in various Bayesian task-based 
% analyses \cite{zhang2015bayesian,mejia2020bayesian}, but these often introduce some confusion on reasonable
% activation criteria. 
The approach we take herein is to assign a spatial 
sparsity prior distribution on $\beta_v$ of the form in 
(\ref{eq:spat_spars_prior}). Here, $N^+(\cdot, \cdot)$ 
denotes the half-normal distribution
and $t_a^+(\cdot, \phi)$ denotes the half-t distribution with $a$
degrees of freedom and scale parameter $\phi$. 
\begin{equation}
  \beta_v | \lambda_v, \tau \sim N^+(0, \lambda_v^2\tau^2); \quad
    \lambda_v|\phi_v \sim t_{\nu}^+(0,\phi_v)
    % \tau &\sim N^+(0, \tau_0^2)  \nonumber
\label{eq:spat_spars_prior}
\end{equation}
The prior distribution in (\ref{eq:spat_spars_prior}) belongs to the 
global-local scale mixture of normal shrinkage 
priors, which are used when coefficients or
signals are sparse. 
% meaning the majority of coefficients are expected to
% be zero. 
The parameter $\lambda_{\nu}$ determines the local shrinkage at voxel $v$ and
the parameter $\tau$ determines the global shrinkage shared by signals at all
voxels. 
% Often a prior distribution is assigned to the global
% shrinkage parameter $\tau$.
The prior distribution in (\ref{eq:spat_spars_prior})
is notably
similar to the
horseshoe (HS) sparsity prior which
was designed to provide spike and slab like behavior but
with a continuous distribution, making computation much simpler
\citep{carvalho2009handling,carvalho2010horseshoe}.
Two features of the HS prior provide it with excellent shrinkage
properties including having an infinite mass at zero
and
heavy tails which allow signals to escape the 
strong pull to zero. We refer to the prior in 
(\ref{eq:spat_spars_prior}) as the half-t horseshoe (HTHS)
prior.

The HTHS prior differs from the HS in 
two important ways, primarily with respect to the
hyperprior distribution on the 
local shrinkage parameter $\lambda_v$.  
First, the use of the half-t prior in place of the 
half-Cauchy provides a simple way to adjust the distribution shape by
changing the degrees of freedom parameter $v$. Increasing $v$ to be 
greater than 1 causes lighter tails and slightly
adjusted spike
behavior at zero, although the infinite spike is 
maintained as shown by Theorem \ref{prop:inf_spike}.
% and illustrated in Figure \ref{fig:marg_priors}. 
% Figure \ref{fig:marg_priors} shows that 
% swapping the half-Cauchy distribution for the half-t and 
% integrating over 
The marginal distribution in Theorem \ref{prop:inf_spike} is the HTHS integrated over $\lambda_v$. 
Without 
loss of generality, we assume $\tau = 1$ and drop the subscripts
$v$ to simplify notation. A proof is given in the 
supplementary materials, along with a figure comparing the marginal
distribution of the HTHS prior with the HS.
% the marginal density
% with $\nu \geq 1$
% maintains the infinite mass at 0. Here, to simplify notation but without
% loss of generality, we assume $\tau = 1$ and drop the subscripts
% $v$. 
\begin{theorem}
\label{prop:inf_spike}
For $\tau^2 = 1$, $\phi \in (0, 1]$ and degrees of freedom $\nu \geq 1$, the density $p(\beta|\phi)$
for the half-t horseshoe prior satisfies
\[
  \lim_{|\beta| \to 0} p (\beta | \phi) = \infty
\]
\end{theorem}
% Under the HTHS prior, 
% shrinkage to 0 for moderate or small signals, while still strong, is weaker than 
% that inherent in the HS. At the same time, the lighter tails from the 
% half-t make it more difficult for strong signals to escape. 
% Figure \ref{fig:marg_priors} compares the marginal 
% HTHS with $\nu = 1000$ with
% the marginal HS prior where $\tau = \phi = 1$ and 
% with the standard normal distribution. The figure shows 
% plots for bulk of the prior distributions near zero as well 
% as the distribution tails. 

% \begin{figure*}[!hbt]
%     \centering
%     \includegraphics[scale=.55]{images/marg_priors.png}
%     \caption{Marginal horseshoe, half-t horseshoe, and 
%     standard normal prior distributions. Here $\tau = 1$ and 
%     $\phi = 1$.}
%     \label{fig:marg_priors}
% \end{figure*}

The additional control on the prior shape 
% from using the half-t in place of the half-Cauchy
is important when 
modeling naturalistic data with model (\ref{eq:general_model})
due to the simultaneous 
estimation of $f(\cdot)$ and $\beta_v$.
% and their
% lack of 
% identifiability. 
As shown in the simulation study in 
Section \ref{sec:analyses}, 
for large degrees of freedom $\nu$ on the local shrinkage parameter
$\lambda_v$ provides
improved shrinkage when there is no true activation
signal as compared to small $\nu$.
The primary benefit 
is to make the tails lighter
and thus making more difficult
that false 
signals escape the shrinkage to zero. 
% This
% added control is important where $\beta_v$ and
% $f(\cdot)$ are estimated together and are not 
% identifiable. The simulation study in the following section
% demonstrates how for small degrees of freedom in the 
% HTHS prior, voxel activation where no true
% activation is present can be falsely detected because
% of the heavy tails on the prior.
% power detecting
% SNR than small $\nu$.
For the work done herein, we prefer $\nu = 1000$ which makes the 
prior nearly a half-normal, though fitting results were very
similar between $\nu = 20$ and $\nu = 1000$. However, we
recommend using finite $\nu$ because
the light tails in the half-normal prior fail to meet 
the conditions established by
\citet{ghosh2015posterior} and
\citet{van2016conditions} for a minimax
optimal posterior contraction rate.

Due to lack of identifiablity of $f(\cdot)$ and $\beta_v$, care must be taken
when assigning a hyperprior to the global parameter
$\tau$. In the 
linear model case with a known predictor, 
\citet{polson2010shrink}
recommend the prior distribution
$\tau \sim C^+(0, \tau_0)$ where $C^+(\cdot,\cdot)$ denotes the
half-Cauchy distribution.
\citet{piironen2017hyperpriora,piironen2017sparsityb}
also recommend the 
half-Cauchy distribution along with prior information to select
the value of $\tau_0$ . 
However, where $f(\cdot)$ in model (\ref{eq:general_model}) must
be estimated, we found that the heavy tails of the half-Cauchy 
led to poor estimation of $f(\cdot)$ and higher active SNR
voxels than should be expected. Thus, regularization on
$\tau$ is desired, and we assign the prior 
distribution $\tau \sim t^+_{1000}(0, \tau_0)$, where the lighter
tails make it more difficult for $\tau$ to be too easily enlarged.
% Piironen and Vehtari
%  \cite{piironen2017hyperpriora} and 
% \cite{piironen2017sparsityb} provide a principled approach for selecting the
% hyperprior on the global shrinkage parameter $\tau$. 
% They use some theory from
% a different paper, and that suggestion is roughly though not 
% exactly followed in this paper. 
Using the framework given by
\citet{piironen2017hyperpriora, piironen2017sparsityb}, 
we set $\tau_0 = \tau^*/\sqrt{S\times V \times T}$,
where $S$ is the number of participants, $V$ is the total
number of voxels, and $T$ is the total number of 
time steps.
We set the value $\tau^* = 0.1$, which may be considered
our rough prior estimate of the
proportion of active voxels in a naturalistic SNR
experiment. As discussed by 
\citet{piironen2017hyperpriora}, activation can be
sensitive to the value of $\tau^*$, but as shown in small exploration of other values in the supplementary material, activation is not too effected for $\tau^* \leq 0.1$. 
Between fixing the 
GP variance parameter to be equal to 1, constraining
$\beta_v$ to be nonnegative, and using the HTHS prior in place of the HS,
the identifiability concern was largely resolved.

To determine whether or not to declare a voxel 
as activated by the naturalistic stimulus,
the posterior shrinkage factor given in 
(\ref{eq:shrink_fact}) is estimated and then compared with
a determined cutoff value.
This shrinkage factor is similar to that given by 
\citet{carvalho2010horseshoe}
and \citet{piironen2017sparsityb}
who declare a coefficient assigned the HS prior to be
significant if $E[\kappa_v | \tau, \sigma, \boldsymbol{f}] < 0.5$.
We estimate $E[\kappa_v | \tau, \sigma, \boldsymbol{f}]$
as the mean of posterior draws of the parameter
$\kappa_v$, and declare a voxel active if that value
is less than 0.5. The cutoff of 0.5
is usually reasonable as the large majority of posterior
means of $\kappa_v$ tend to be very close to 0 or 1
signaling a clear escape of the coefficient away from 
0 or strongly maintained shrinkage to 0 respectively. 
This is demonstrated in the simulation study in the 
following section.
\begin{equation}
    \kappa_v = \frac{1}{1 + S \sigma_{\epsilon}^{-2} \tau^2 \lambda_v^2 \boldsymbol{f}^{\top} \boldsymbol{f}}
    \label{eq:shrink_fact}
\end{equation}
% \cite{piironen2017sparsityb}
% Another modification is to use
% here a half-normal distribution instead of a half-Cauchy distribution. 
The HTHS prior on the activation coefficient $\beta_v$ 
treats voxel activation as 
conditionally independent, but as activation is expected to be similar among
neighboring voxels, a spatial component is added to the local
shrinkage parameter. This is done by modeling the
scale parameters of the prior on $\lambda_v$
$\boldsymbol{\phi} = (\phi_1, ..., \phi_V)$ as a latent spatial 
field which follows an intrinsic Gaussian Markov 
random field (IGMRF) model defined in (\ref{eq:icar1}).
Here, $N(v)$ is the 
set of neighbors to voxel $v$ --not including voxel $v$-- and
$d_{N(v)}$ is the number of neighbors. A neighbor
of voxel $v$ is defined as a voxel which shares a face 
$v$ which in three dimensions makes the 
maximum number of total neighbors six. 
\begin{equation}
  \begin{aligned}
    \phi_v &= \text{logit} (\alpha_v) \\
    \alpha_v | \alpha_{N(v)} &\sim
            N\left(\frac{\sum_{\alpha_i \in N(v)} \alpha_i}{d_{N(v)}},
                   \frac{1}{d_{N(v)}}\right)
  \end{aligned}
  \label{eq:icar1}
\end{equation}

The IGMRF model, 
introduced by \citet{besag1991bayesian}, assumes a high level of correlation
amongst neighbors and is a 
case of a Gauss Markov random field (GMFR) model.
GMRF models are characterized by their precision
matrices which are made sparse by assuming conditional independencies.
This sparsity leads to fast computation while still allowing for 
reasonable correlation structures
\citep{rue2005gaussian}. 
The IGMRF model is an improper distribution, but when 
combined with a likelihood the posterior distribution often is proper,
and a sum to zero constraint $\sum_i \alpha_i = 0$ ensures 
propriety \citep{lee2011comparison}.
The spatial smoothing applied to the local shrinkage 
parameters encourages neighboring voxels to show similar
activation. 
% That is, whether or not a voxel displays
% shared neural activity is highly related to whether or 
% not its neighbors do.

Incorporating spatial information into the HTHS prior
adds a novel contribution to the growing literature on
dependent variable selection priors.
\citet{jhuang2019spatial} propose a spatial horseshoe prior
where the prior on $\lambda_v$ is replaced with a Gaussian copula model that 
preserves the half-Cauchy distribution while also accounting for spatial 
dependence. \citet{chang2018scalable}
and \citet{li2024accounting}
introduced a shrinkage prior
which assigns 
independent Laplace priors to regression 
coefficients and incorporates
variable dependence
by modeling the Laplace shrinkage parameters as a
connected graph based on underlying information. \citet{denis2023graph}
extend the work of \citet{faulkner2017locally}
and introduce 
a HS GMRF which encodes 
graph information in local shrinkage parameters of 
priors on coefficient differences.

% The activation of a voxel will naturally be related to the surrounding voxels,
% so it makes sense to borrow spatial information. This is accomplished 
% by modeling the 
% scale parameter of the prior on $\lambda_v$ according to a spatial process. 
% Because of its speed in fitting, we model $\boldsymbol{\alpha}$ according to
% an intrinsic conditional autoregressive (ICAR) model \cite{besag1974spatial}.
% I use four neighbors, which is called the nearest neighbor approach 
% \cite{besag1974}.

% \begin{equation}
%   p(\boldsymbol{\alpha}) \propto \text{exp} \left(-\frac{1}{2} \sum _{v \sim v'}
%       (\alpha_v - \alpha_{v'})^2 \right)
%   \label{eq:icar1}
% \end{equation}

\subsection{BNR Model Fitting}

When fitting model (\ref{eq:general_model}) to whole brain
data from multiple participants as done in the real data
examples in Section \ref{sec:analyses}, the model is 
fit to a select set of regions of interest (ROIs) each independently for two
reasons. First, it is assumed that voxel behavior within
one region is very similar among all voxels in that 
region but that the overall behavior can differ from 
that in other ROIs. Fitting the ROIs separately allows
$f(\cdot)$ to vary by region but also maintains the 
assumption of similar within region behavior.
The second reason for modeling each independently is 
computation. Fitting this way allows for ROIs to be fit
in parallel, greatly improving fit time and scalability.
An obvious drawback is the lack of shared information 
among ROIs and the possibility of starkly different
model behavior of neighboring voxels in differing
ROIs. However, our results did not suggest that these issues 
were very severe, in line with previous findings in the 
literature \citep{musgrove2016fast,wang2024fully}.

The model is fit via Hamiltonian Monte Carlo sampling 
using the \texttt{cmdstanr} package in \texttt{R},
which employs the No-U-Turn sampler
\citep{hoffman2014no, gabry2022stan, stan2024manual}.
The NNGP is evaluated using the 
\texttt{gptools} package which interfaces with 
\texttt{Stan} \citep{hoffmann2025gptools}.
For voxel activation,
% $E[\kappa_v | \tau, \sigma, \boldsymbol{f}]$ is estimated
% as the mean of posterior draws of $\kappa_v$.
To assess posterior convergence, we fit the BNR model to 
data in an ROI with several detected active voxels by
drawing 3000 posterior samples from four chains --dropping the first 
1500 as a burn-in-- and checking
diagnostic statistics. Among all parameters, the 
maximum $\hat{R}$ statistic was 1.006, the minimum
effective sample size was 1,899, and the minimum tail
effective sample size was 1,211. The recommended values to
meet for these are a maximum $\hat{R}$ below 1.01 and a few
hundred effective samples.
The intended use for the BNR model is for nfMRI experiments
involving up to dozens of participants, a typical sample siz.
Adjustments to the fitting mechanism such as using
fitting via integrated nested Laplace approximation rather than
MCMC sampling
could produce speedups which allow larger sample
sizes \citep{zhang2015bayesian,mejia2020bayesian}, the cost 
being a potential reduction in inference quality.

\section{Data Analyses} \label{sec:analyses}
% \spencer{
% Possible items to add
% \begin{itemize}
%     \item Discussion on posterior of $\kappa$ with images in supplementary 
%     materials
% \end{itemize}
% }
In this section, the BNR model's ability to predict SNR activation and 
estimate a SNR function
is assessed by fitting the model to simulated nfMRI data,
real tfMRI data, and real nfMRI data. The results are compared with 
standard methods used for these analyses.

\subsection{Simulation Study} 

The data for this study was simulated using the 
\texttt{neuRosim} package in \texttt{R} \citep{welvaert2011neurosim}. 
The fMRI data is simulated from model (\ref{eq:bold_glm}), where
the user 
defines the forms of $h_k(t)$ by specifying a task design --including the
duration of the task and the task onset times-- and convolving the design
with the hemodynamic response function. 
The intention was to make a complicated BOLD signal, so
four separate tasks were specified, each having its 
own onset times and task durations. The total time was set to 240 seconds,
and TR was set to 2 making in total 120 temporal observations per participant.
The study was done under 12 different settings including four BOLD
signal strength levels and three different number of participants 
$S \in \{5, 10, 20\}$. 
The BOLD signal strengths for all tasks were
$\beta_{k,s,v} \sim FN(\mu, \sigma^2)$ for 
$\sigma = 0.34$ and $\mu \in \{0.5, 1, 2\}$ representing a weak, medium, or strong signal or $\beta_{k,s,v} = 0$ for no signal.
$FN$ denotes the folded normal distribution which 
was used to ensure $\beta_{k,s,v}$ was positive. 
On an $11\times 11$ voxel grid, a circular activation 
region was specified
where the maximum signal was in the center 
of the circle. For each participant the circle center was the same,
but the radius was drawn from the normal distribution
$N(2.5, 0.56^2)$. The maximum signal was at the center 
of the
circle with a slight fading to the edge. 
% The top row of Figure \ref{fig:sim_design} shows the centered and scaled
% average shared response
% signal along with the activated voxels on the voxel grid.
% \begin{figure*}%[!hbt]
%     \centering
%     \includegraphics[width=0.7\linewidth]{images/sim_design.png}
%     \caption{(a) Specified shared response function. (b) Example activation
%     region coloured by relative signal strength. (c) Example simulated 
%     BOLD response from 20 simulated participants at an active voxel. (d)
%     Example BOLD response at one timepoint over $11\times11$ grid.}
%     \label{fig:sim_design}
% \end{figure*}
The term $\epsilon_{s,v}(t)$ in model (\ref{eq:bold_glm}) 
captures noise and possible nuisance 
variables. The \texttt{neuRosim} package allows for simulating 
various types
of noise, including spatial, temporal, physiological, low-frequency, and
white noise. We specified the AR(3) temporal process
\[
\gamma_s(t) = \eta_1\gamma_s(t-1) + \eta_1\gamma_s(t - 2) + \eta_1\gamma_s(t - 3) + \xi(t)
\]
where $\boldsymbol{\eta} = (\eta_1, \eta_2, \eta_3) \sim MVN((0.142, 0.108, 0.084), 0.3\boldsymbol{I})$
and $\xi(t)$ is i.i.d normal for all $t$.
The remaining spatial,
physiological, low-frequency and white noise were set to default
settings except that the white noise was Rician distributed 
based on a low selected signal-to-noise ratio and previous
research on fMRI noise \citep{welvaert2011neurosim, adrian2025rice}.

\begin{figure*}[!hbt]
    \centering
    \includegraphics[width=0.55\linewidth]{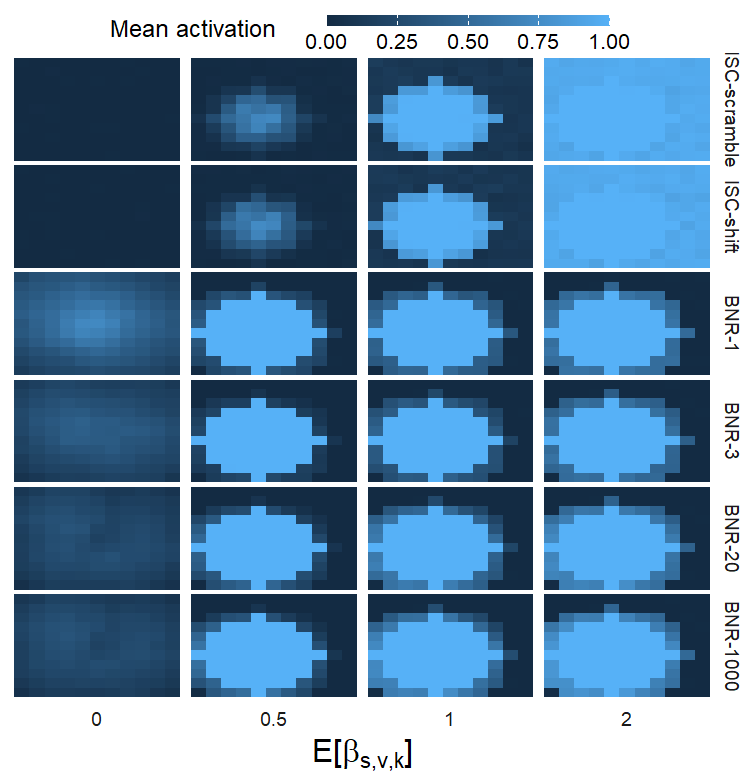}
    \caption{Mean over 200 simulated replicates of shared response activation predicted at each voxel on the $11\times11$ grid. Columns are separated by signal strength from none to high, and rows separate results by activation prediction model.}
    \label{fig:tile_activation}
\end{figure*}

\begin{figure*}[!hbt]
    \centering
    \includegraphics[width=.9\linewidth]{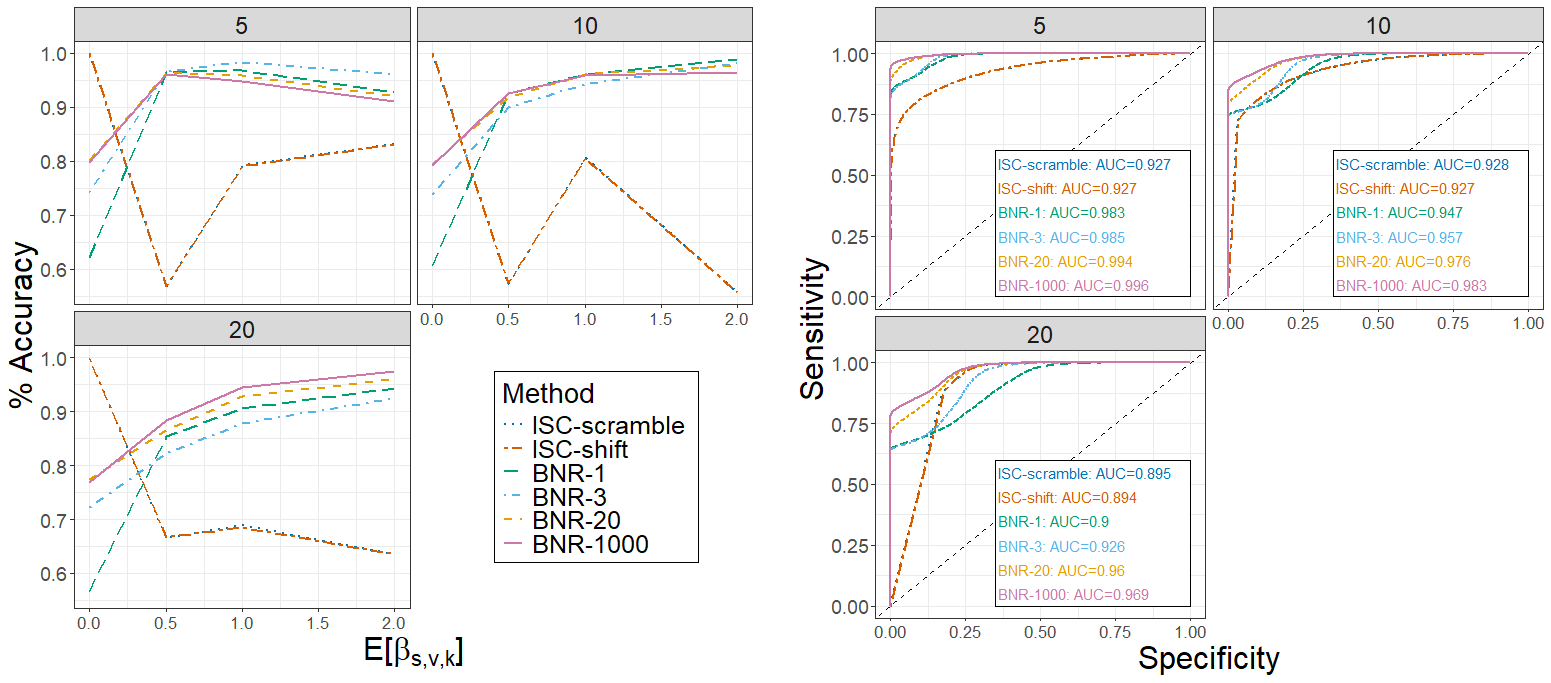}
    \caption{(left) Overall classification accuracy 
    score of active voxels 
    number of participants. Each plot shows accuracy for
    increasing signal strength, and scores are 
    separated by 
    classification model. (right) Receiver operating
    characteristic (ROC) curves and area under the 
    curve (AUC) for each activation detection method 
    separated by number of participants.}
    \label{fig:snr_roc}
\end{figure*}

\begin{figure*}[!hbt]
    \centering
    % \includegraphics[width=0.45\linewidth]{images/f1_bar.png}
    % \hspace{.4cm}
    \includegraphics[width=0.45\linewidth]{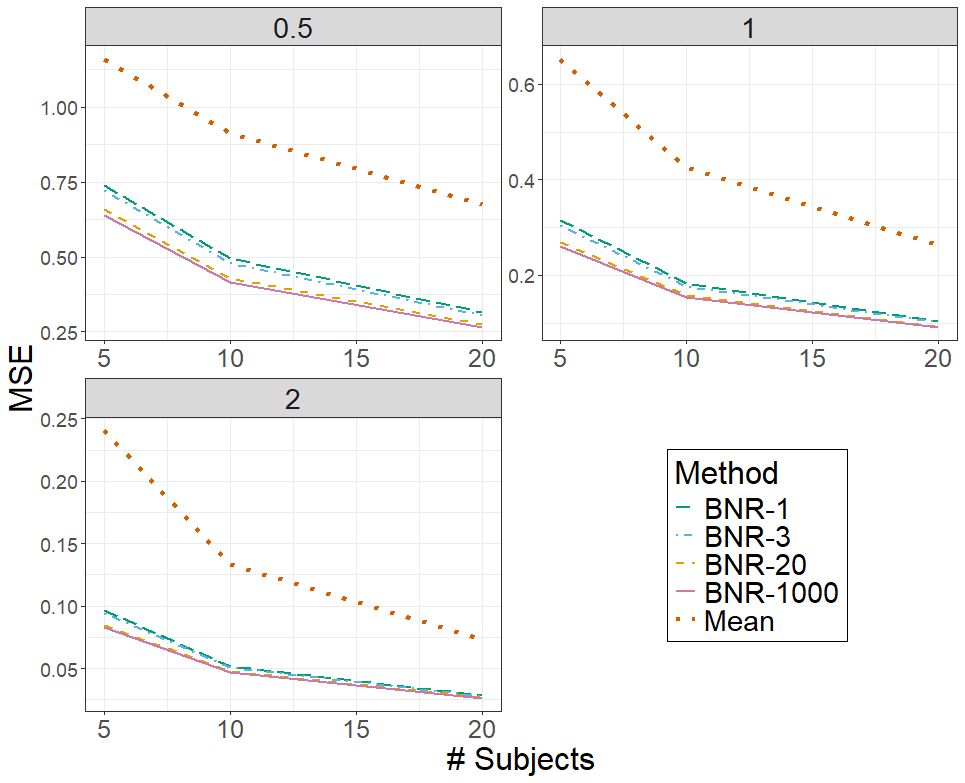}
    \caption{Average of 200 simulation replicates
    mean square error (MSE) between the true
    shared neural response function 
    and the estimated function coloured by model or by mean.}
    \label{fig:snr_mse}
\end{figure*}

% The bottom row of Figure \ref{fig:sim_design} includes a plot
% with centered
% and scaled simulated fMRI data from 20 participants at an activated 
% voxel. It also includes a plot of the BOLD value on the grid at a fixed
% time to show the added spatial noise.
For each
of the 12 settings, 200 replicates were simulated to which models were
fit for estimating voxel activation and the SNR function.
Before fitting, the data was centered and scaled to have standard deviation 1.
Figures \ref{fig:tile_activation},
\ref{fig:snr_roc} and \ref{fig:snr_mse} show the results of fitting the BNR-$\nu$ model to the
simulated data for differing degrees of freedom 
$\nu \in \{1, 3, 20, 1000\}$
on the HTHS prior. The results for activation and estimation are
compared with activation estimated using ISC where sampling distributions
for testing were estimated using circular shifting or 
phase scrambling resampling methods. 

In Figure \ref{fig:tile_activation}, the average voxel activation over all
200 simulation replicates is shown under different signal strength values
for number of participants $S = 10$. 
The ISC methods show low statistical power for 
a weak BOLD signal, and they have a high false positive rate when the
signal is strong, up to 30\%. They do well at predicting voxel 
activation for medium signals and almost never predict 
false positive activation when there is no signal. 
All BNR models improve over the ISC methods in predicting
voxel activation for weak to strong signals and have low false
positive rates up 4\%. When the signal is 
zero, they tend to predict some false
positive active voxels. However, as the degrees of 
freedom $\nu$ on the HTHS prior increases, the tendency
to falsely predict active voxels decreases significantly from 
40\% when $\nu = 1$ to 21\% when $\nu = 1000$.
Overall, the BNR models with higher $\nu= 1000$ performs
the best at detecting voxels that are
active where the SR signal is non-zero and minimizing 
the false positive activation prediction.

The plots on the left of Figure \ref{fig:snr_roc} show the overall  
classification accuracy
for each model under different levels of
signal strength. 
For all signal strengths, the BNR models tend to have 
similar accuracy with $\nu \in \{20, 1000\}$ slightly
outperforming the lower degrees of freedom models. 
The BNR models outperform ISC for all signal strengths 
except where the signal is zero. The plots on the right
of Figure \ref{fig:snr_roc} show the receiver
operating characteristic (ROC)
curves for the BNR and ISC methods. The ROC curves, 
along with the area under the curve (AUC) values show 
that the BNR models with $\nu \in \{20, 1000\}$ 
achieve the highest sensitivity and specificity. Plots of
the posterior expected values of the shrinkage factor
$\kappa_v$ are included in the supplementary materials
along with an additional brief 
discussion of voxel activation under different prior distributions.

The plots in Figure \ref{fig:snr_mse} show the mean square
error (MSE) between the true SNR function and the function estimated by the 
model or by estimating the function as the mean over time points from the
data. The BNR-$\nu$ models clearly outperform
estimating the SNR function by taking the mean over all
participants,
with the larger $\nu$ models performing the best.
Overall, 
the BNR-$\nu$
models with large $\nu$ perform the best at both classification and 
function estimation. One drawback is the tendency to classify some voxels as
active when no signal is present.

% \begin{figure}[!hbt]
%     \centering
%     \includegraphics[width=0.5\linewidth]{images/mse_line.png}
%     \caption{Caption}
%     \label{fig:mse_score}
% \end{figure}

\subsection{Task-based Checkerboard Study}

An important assessment of the BNR model's ability to detect shared activation
and estimate a SNR function from real data is to 
compare its performance against that of a GLM
fitted to fMRI data from
a simple task-based experiment. Given a well designed task-based experiment,
we would not expect the BNR model to outperform the GLM, but it should be 
able to create a similar activation map as well as estimate a shared 
response function with a trajectory similar to the BOLD predictor used in
the GLM. 
% A similar activation profile and an accurately estimated SNR function would indicate that the BNR model may reasonably be used
% for analyses of more complicated naturalistic data.
In this study, we fit the GLM and BNR models
to the same t-fMRI dataset --involving a single task-- 
and we compare the activation profiles of the two
models. We also compare the BNR's estimated shared neural response functions
to the BOLD predictor used in the GLM.
The data in this study comes from  
\citet{telesford2023open}.

\citet{telesford2023open}
collected fMRI data on 22 adult participants who
were each presented with several different stimuli, each on a different
scanning occasion. One of these was a
checkerboard stimulus. The checkerboard stimulus involved showing each 
participant a radial checkerboard which flickered at 12 Hz for a 20 second period. Following the viewing of the checkerboard, a neutral rest
image was shown for 20 seconds, and the task was repeated 
several times. 
The repetition time (TR) was 2.1 seconds, and each voxel
dimension was 
$3.469\times3.469\times3.330$mm.
The data underwent several standard preprocessing steps, including 
registration of scans from each participant to a common template space,
processing
with and without spatial and temporal filtering, and 
processing with and without
global signal regression as well as several other steps, 
see the \emph{Methods} section in 
\citet{telesford2023open}.
Herein, we use the preprocessed data which did not undergo filtering but which
did undergo global signal regression.
The data was then separated into 400 
ROIs from the 
Schaefer 400 atlas \citep{schaefer2018local}. The number of voxels 
per ROI ranged from 4 to 247 with the median number of voxels being 78.

For the GLM, model (\ref{eq:bold_glm}) 
was fit where $K = 1$ and a common $\beta_{1,v}$ (dropping the
subscript $s$) was estimated. The 
BOLD predictor $h_1(\cdot)$ was made by convolving the task 
block design
with the canonical HDR function. The prior distribution
on the activation coefficient $\beta_{1, v}$ is the
HTHS prior (\ref{eq:spat_spars_prior}) used in the BNR model except 
that the normal distribution was used instead of the half-normal and 
the degrees of freedom was $\nu = 1$. 
This is the standard HS prior.
The prior distribution assigned to the global shrinkage parameter
was $\tau \sim C^+(0, 0.1)$.
For the BNR model the degrees of freedom
on the local shrinkage prior
was $\nu = 1000$, and the prior assigned to the 
global shrinkage parameter was
$\tau \sim t_{1000}^+(0, 0.1)$. Both models were fit to 
each of the 400 ROIs separately. 

\begin{figure*}[!hbt]
    \centering
    \includegraphics[width=0.55\linewidth]{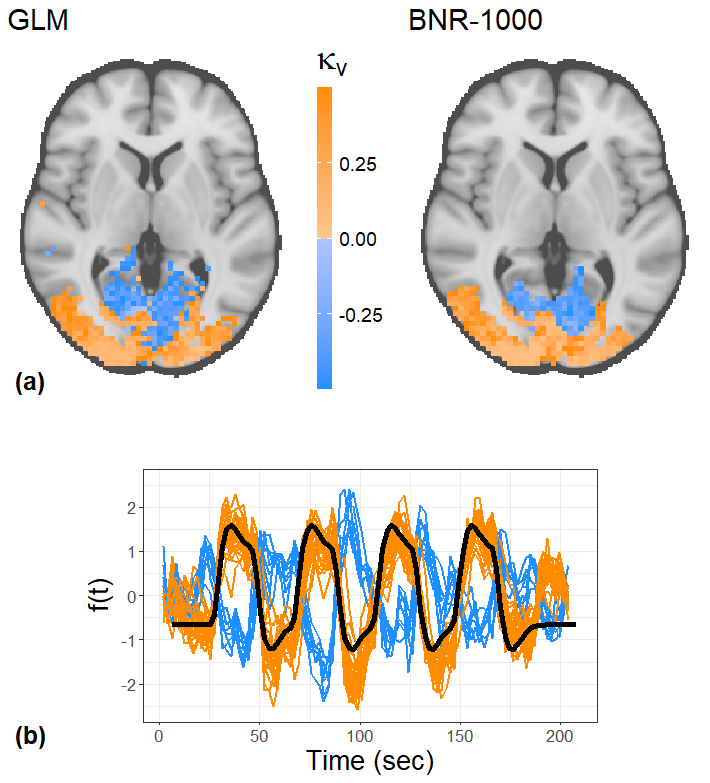}
    \caption{(a) Shared neural activation maps of checkerboard task fMRI
    data for GLM and BNR models. Orange voxels show positive BOLD 
    activation and blue voxels show negative BOLD activation.
    (b) Estimated shared neural response function from BNR model for 
    ROIs where any number of voxels was predicted as active. Functions 
    are overlaid with GLM BOLD predictor function for the experiment 
    design.}
    \label{fig:check_act}
\end{figure*}

Figure \ref{fig:check_act} shows SNR activation results for the two models as well as the 
BNR predicted SNR function for ROIs which contained active voxels.
In the activation maps, the orange shows activation for
the BOLD response, whereas the blue shows activation for
the negative BOLD response. In the case of the GLM
negative bold response means the mean estimated 
$\beta_{1,v} < 0$, and for the BNR the estimated 
$f(\cdot)$ followed a negative trajectory of 
$h_1(\cdot)$.

The activation maps 
in plot (a) from the two models are very similar
with the BNR model displaying slightly stronger
spatial relationships in that some of the voxels declared
active by the GLM model have few or no active neighbors. 
It also reflects the fact that the BNR model does not 
allow for a single ROI to contain active voxels of both
BOLD and negative BOLD response. 
Plot (b) shows the mean predicted shared neural 
response functions for the ROIs with any number of
active voxels under the BNR model. The predicted functions
are overlaid with the BOLD predictor $h_1(t)$ used in 
the GLM. The predicted functions clearly show a
pattern similar to the designed predictor, suggesting the
BNR GP does a reasonable job estimating a true shared neural
response.

% \begin{figure*}[!hbt]
%     \centering
%     \includegraphics[width=0.5\linewidth]{images/check_act_vox_count.png}
%     \caption{Histograms showing count of number of 
%     predicted activated voxels per
%     ROI for GLM and BNR models on checkerboard fMRI data.}
%     \label{fig:check_act_vox_hist}
% \end{figure*}

Of the 400 ROIs, the GLM predicted at least one activated
voxel in 80 and the BNR predicted at least one activated
voxel in 35. However, of the 80 ROIs from the GLM, 26 
of these contained only one activated voxel. 
% Figure \ref{fig:check_act_vox_hist} shows 
% histograms of the activated voxel count per ROI. 
Additional plots in the supplemental materials are shown to 
compare BNR-1000 fits with those of the 
BNR-1, BNR-3, and BNR-20 models. 
The results show that the BNR-1 and BNR-3 models 
predict far more active voxels
than the GLM. This suggests that 
either there is far more shared neural response than can 
be detected using the GLM, or more likely that the BNR-1 and
BNR-3 models make spurious predictions.

% This comparison in this study of the GLM and BNR model 
% analyses suggests the BNR is capable of 
% accurately predicting a shared neural response function
% and voxel activation in task-fMRI settings. 

\subsection{Naturalistic Stimulus Study}

The study in this section demonstrates the BNR model fit to naturalistic
fMRI data.
The data in this study comes from \citet{telesford2023open}
and 
belongs to the same dataset which was used
in the previous study. Besides the checkerboard stimulus,
the participants were presented with 
several different naturalistic videos. 
The nfMRI data here
were collected from the same 22 adult
participants while they viewed a 10 minute clip of the 
movie \emph{Despicable Me}. The exact playing times
of the clip are 1:02:09 - 1:12:09. The data underwent the same preprocessing
steps as the checkerboard data, and it was separated into the same 
400 ROIs and 97 time steps from the original data were used. 
The BNR-3 and BNR-1000 models were fit to the data for 
predicting voxel activation, and ISC voxel activation prediction 
was also performed on the data under both 
circular shifting and phase scrambling methods. 

\begin{figure}[!hbt]
\centering
% \begin{figure*}[!hbt]
\begin{subfigure}[c]{0.35\textwidth}
    % \centering
    \caption{}
    \includegraphics[width=0.9\linewidth]{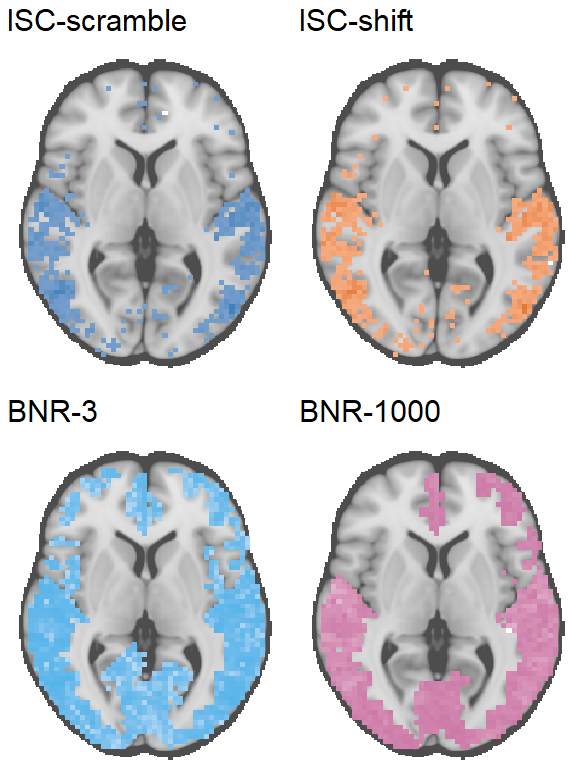}
    % \caption{Shared neural activation maps of naturalistic fMRI data 
    % where participants viewed a clip from \textit{Despicable Me}. Results
    % are show for
    % ISC circular shift and phase scrambling methods
    % and two BNR models.}
    % \label{fig:dme_act}
% \end{figure*}
\end{subfigure}
% \begin{figure}[!hbt]
    % \centering
\begin{subfigure}[c]{0.6\textwidth}
\centering
    \caption{}\
    \vspace{.8cm}
    \includegraphics[width=1\linewidth]{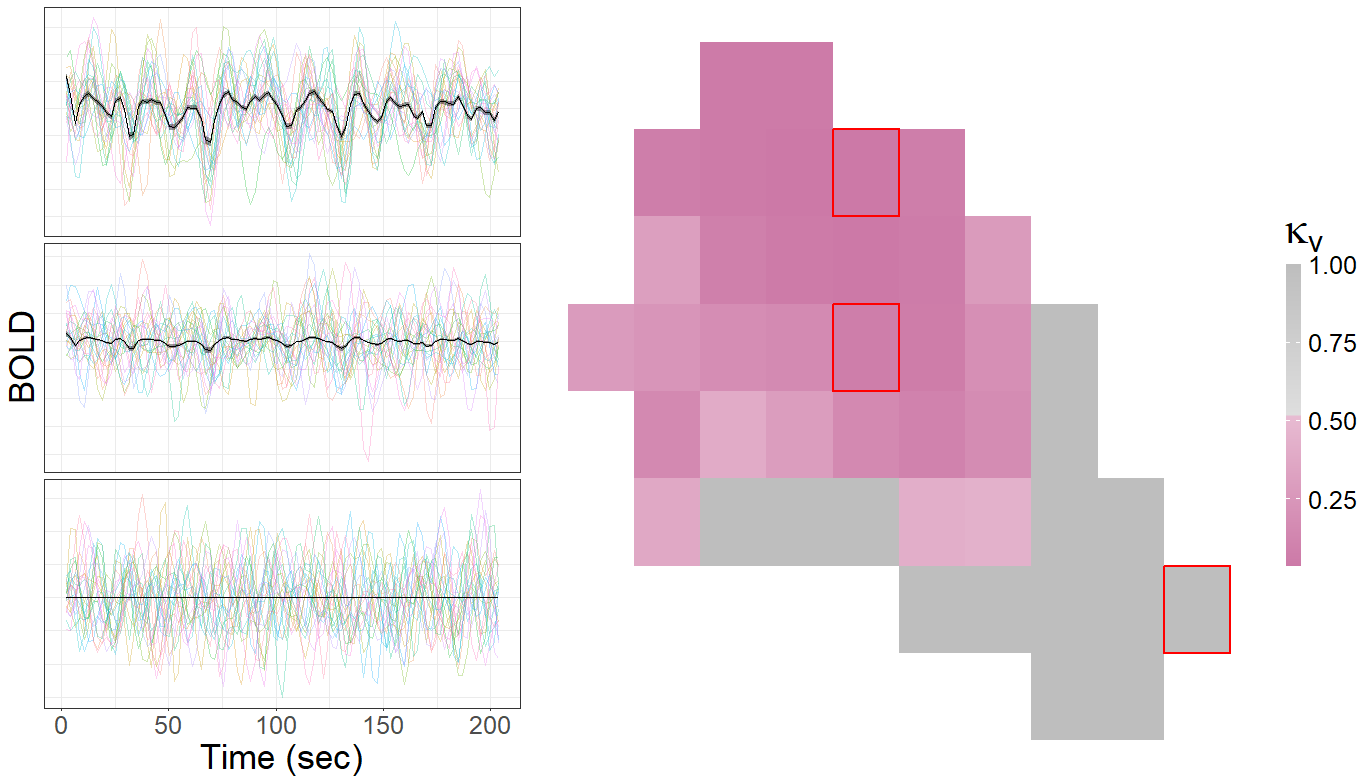}
\end{subfigure}
    \caption{(a) Shared neural activation maps of naturalistic fMRI data 
    where participants viewed a clip from \textit{Despicable Me}. Results
    are show for
    ISC circular shift and phase scrambling methods
    and two BNR models.\\
    (b)
    (left) Neural response data from 22 participants during 
    viewing of Despicable Me overlaid with 
    estimated shared neural response function at three 
    voxels with 
    varying levels of estimated activation by the BNR-1000 model. 
    (right) Activation map for one ROI with outlined voxels corresponding from
    top to bottom to the three estimated function on the left.}
    \label{fig:dme_est}
% \end{figure}
\end{figure}

Part (a) of Figure \ref{fig:dme_est} shows the 
SNR activation maps for the four methods. Both BNR activation maps 
predict voxel activation in roughly 
the same areas as the ISC models, but 
the BNR activation maps show a much higher spatial relationship 
among neighboring voxels. The BNR-3 shows much higher activation
than the BNR-1000, though the BNR-3's activation is likely higher
than should be expected.
Particularly notable is that the BNR models show higher
activation
in the visual cortex than the ISC, which is expected to be highly
active during a visual stimulus \citep{di2023individual}.
Also notable is the BNR-1000 displays much higher activation on the
right side of the prefrontal cortex than the left 
which has been noted in other studies as characteristic 
during naturalistic viewing tasks \citep{jaaskelainen2008inter}. 
Part (b) of Figure \ref{fig:dme_est} shows estimated SNR
functions at three different voxels with 
varying strengths of activity. The estimated functions where activity is 
estimated as stronger appear to follow trends from the BOLD time series of
the 22 participants. Note that participant data shown in 
Figure \ref{fig:dme_est} is spatially filtered to help visualize that the 
estimated SNR function appears to fit well the true real data signal. 
The data used for fitting, however, was unfiltered, showcasing the 
BNR's ability to capture the true SNR signal.

\section{Conclusion} \label{sec:conclusion}

In this manuscript, we introduced a Bayesian modeling framework
for simultaneously estimating a fMRI SNR function and detecting voxel
activation of multiple participants exposed to a naturalistic stimulus.
The modeling framework is a generalization of the classical GLM used
in tfMRI studies, where rather than supplying functional predictors
the SNR function is estimated given the data.
The model relies on sparse Gaussian random field modeling to assist
in making more rapid sampling from the 
posterior distribution.
The combined estimation of the SNR function and the voxel activation
map require thoughtful selection of the prior distribution
on activation coefficients. This led to the development and
study of the half-t generalization of the horseshoe prior.
In a simulation study, the model outperformed standard ISC methodology
for analyzing nfMRI data, and in a real data analysis it 
performed similar to the classical GLM on tfMRI data.
In a real data analysis of nfMRI data, the model detected shared 
responses in areas of the brain considered inactive using ISC
methods, and produced a smoother spatial activation map.

In a clinical setting, a natural extension to the model 
used herein would be to incorporate group information in 
the spatial activation and/or the estimated SNR 
function. 
A multi-group spatial field or GP could be used to capture 
group differences and allow for coherent analysis where 
group responses would be expected to be different, e.g. where
one group of participants is diagnosed with a particular condition or
disease and
the other group is categorized neurotypical.
As of the most recent writing
of this manuscript, the authors were exploring methodology
addressing the two previously mentioned lines of inquiry.

Time for model fitting is a major bottleneck, and certain decisions
made by the authors to handle this 
can be improved upon in future work. 
For instance the decision to fit each ROI independently was 
necessary and not unreasonable, but jointly modeling all regions 
would likely lead to improved fits particularly on the edges of 
the regions. A potential avenue for accelerating the time to fit 
would be to approximate the posterior distribution via integrated
nested Laplace approximation. Laplace approximation has been
used successfully on tfMRI data, and it provides a good framework
for modeling non-Euclidean cortical surface data \citep{mejia2020bayesian, siden2021spatial}. The methodology herein was developed for small
sample sizes (dozens of participants), and scaling to larger datasets of 
hundreds or thousands of participants will demand more efficient and 
scalable model fitting.

% \section{Appendix}